\documentclass[nohyper,twoside]{JHEP3}
\usepackage{amsmath}
\usepackage{amscd}
\usepackage{amssymb}
\usepackage{array}

\newcommand{\rtr}{\mathrm{tr}}

\newcommand{\rUSp}{\mathrm{USp}}
\newcommand{\rSU}{\mathrm{SU}}
\newcommand{\rE}{\mathrm{E}}
\newcommand{\rSO}{\mathrm{SO}}
\newcommand{\rSL}{\mathrm{SL}}
\newcommand{\rGL}{\mathrm{GL}}

\newcommand{\fsl}{\mathfrak{sl}}
\newcommand{\fgl}{\mathfrak{gl}}
\newcommand{\fe}{\mathfrak{e}}

\newcommand{\fsolv}{\mathfrak{solv}}
\newcommand{\fso}{\mathfrak{so}}

\def\R{\mathbb{R}}

\def\L{\mathbb{L}}

\newcommand{\cV}{\mathcal{V}}
\newcommand{\cG}{\mathcal{G}}
\def\cM{\mathcal{M}}

\preprint{CERN-PH-TH/2005-022\\FTUV-05/0208\\IFIC/05-14}

\title{The Scherk--Schwarz mechanism as a flux compactification with
internal torsion}
\author{
L. Andrianopoli\\
Centro E. Fermi, Compendio Viminale,  I-00184 Rome, Italy\\
E-mail: \email{Laura.Andrianopoli@cern.ch}}
\author{
M. A. Lled\'o
\\
 Departament de F\'{\i}sica Te\`orica,
Universitat de Val\`encia and IFIC,
 C/Dr.
Moliner, 50, E-46100 Burjassot (Val\`encia), Spain.\\
E-mail: \email{Maria.Lledo@ific.uv.es}}
\author{
M. Trigiante\\  Dipartimento di Fisica, Politecnico di Torino,

 C.so Duca degli Abruzzi,  24 I-10129 Torino, Italy\\
E-mail: \email{Mario.Trigiante@to.infn.it}}

\abstract{The aim of this paper is to make progress in the
understanding of the Scherk--Schwarz dimensional reduction in
terms of a compactification in the presence of background fluxes
and torsion. From the  eleven dimensional supergravity point of
view, we find that a general $\rE_{6(6)}$ S-S phase may be
obtained by turning on an appropriate background torsion, together
with suitable fluxes, some of which can be directly identified
with certain components of the four-form field-strength.
Furthermore, we introduce a novel (four dimensional) approach to
the study of dualities between flux/torsion compactifications of
Type II/M--theory. This approach defines the action that duality
should have on the background quantities, in order for the
$\rE_{7(7)}$ invariance of the field equations and Bianchi
identities to be restored also in the presence of fluxes/torsion.
This analysis further implies the interpretation of the torsion
flux as the {\em T-dual} of the NS three-form flux.}

\begin{document}

\def\um{{\underline{m}}}
\def\un{{\underline{n}}}
\def\up{{\underline{p}}}
\def\uq{{\underline{q}}}
\def\ui{{\underline{i}}}
\def\uj{{\underline{j}}}
\def\uk{{\underline{k}}}
\def\ul{{\underline{\ell}}}

\def\deom{{\Delta\Omega}}
\def\cv{\mathcal{V}}

\def\id{\relax{\rm 1\kern-.35em 1}}

\def\eq#1{(\ref{#1})}

\section{Introduction}
This note is intended to shed light on the interpretation of the
Scherk--Schwarz mechanism for generalized dimensional reduction
\cite{ss} within the framework of string compactifications in the
presence of background fluxes and a non-trivial background torsion
in the internal manifold.

Recently there has been considerable interest in the construction
of phenomenologically viable string models. The role of a warped
metric in generating a large hierarchy of scales, fixed in terms
of charges determined by RR and NS background fluxes, has been
extensively studied in the literature \cite{flux1,km}. A
particular attention has been devoted to microscopic settings
leading to a low energy action with a zero vacuum energy. Models
of this kind are generalized no--scale models
\cite{noscale1,noscale2} which are typically related to gaugings
of suitable non--semisimple global symmetry groups of the
Lagrangian (\emph{flat gaugings}) \cite{ssgauging0,ssgauging1}.
They have been obtained from flux compactifications, mostly in
Type IIB theory by switching on appropriate RR and NS three-form
fluxes in the internal directions \cite{flux2,flux3,fluxss} (fewer
examples were obtained on the Type IIA front
\cite{IIAflux,zwirner}), as well as from generalized dimensional
reduction \`a la Scherk--Schwarz (S-S)
\cite{ss,css,km,ssgauging1,kstt,daf,ssgauging2} of eleven dimensional
supergravity (or any truncation thereof). This is a generalized
type of dimensional reduction on tori, which induces non abelian
couplings and a positive definite scalar potential. Generally, it
yields to spontaneous supersymmetry breaking. The simplest example
is a model describing a single complex scalar field $\phi(x,y)$ on
$\R^{3,1}\times S^1$, with Lagrangian
$${\mathcal L}= \frac 12 \partial _{\hat \mu}\phi \partial ^{\hat
\mu}\phi^* ;\quad
 \hat\mu = (\mu,4);\quad \mu = 0,\dots ,3;\, x^4=y.$$
${\mathcal L}$ is invariant under a global $U(1)$: $ \phi \to {\rm
e}^{{\rm i}\alpha}\phi$, and this allows a generalized
compactification ansatz:
$$\phi(x,y) = {\rm e}^{{\rm i} my} \sum_{n=-\infty}^{+\infty}
\phi_n(x) {\rm e}^{{\rm i} ny/2\pi R}.$$ We note that the
multivaluedness of $\phi(x,y)$  on $S^1$ does not pose problems in
the definition of the lower dimensional theory, since, being the
phase  ${\rm e}^{{\rm i}2\pi mR}$ a  global symmetry of the original
theory, it finally cancels in the Lagrangian. The only effect of the
phase is to shift the four dimensional mass spectrum, so that the
zero-mode has mass $m$. This mechanism may be applied to the case of
$D=5$, $N=8$ supergravity \cite{css}, whose scalars span the manifold
$$\frac{ \rE_{6(6)}}{\rUSp(8)} \subset \frac{\rE_{7(7)}}{\rSU(8)}.$$
The S-S dimensional reduction can be performed by using, as global
symmetry, any subgroup of the $78$--dimensional global symmetry
group $\rE_{6(6)}$. We shall call \emph{S-S generator} a generator
of the  global symmetry transformation entering  the S-S ansatz.
It is known that only if the S-S generators are compact, the
resulting no--scale model admits a (Minkowski) vacuum. For any
other choice, it can be shown that the corresponding no--scale
scalar potential is of run--away type.

As the $\rE_{6(6)}$ global symmetry of the $N=8,\,D=5$ supergravity
is manifest when the theory is obtained by dimensional reduction on
a torus of eleven dimensional supergravity, this pattern should
naturally be embedded in the framework of M-theory, or of Type IIA
supergravity. There is an extensive literature on the subject,
 which includes Refs. \cite{zwirner,strss,sertra}.

However a precise interpretation of the S-S phases in terms of
microscopic (stringy) objects has not been completely understood,
since it seems to include non perturbative stringy degrees of
freedom.

On the other hand, as mentioned above, the pattern of flux
compactification has proven to be  successful in providing viable
models with flat gaugings mostly for Type IIB theory. Some of
these models have then been reinterpreted in terms of Type IIA
theory by exploiting T-duality, although the proper generalization
of this correspondence to vacua in which background fluxes are
present is not yet thoroughly understood. For instance it was soon
apparent that the T-dual of the NS three-form flux in Type II
theory could not, in general, be found among the fluxes associated
to the ten-dimensional p-forms of the string spectrum. There is
strong evidence that a non-vanishing torsion in the background
geometry of the internal manifold should play an important role in
filling this gap and thus completing the duality picture
connecting different superstring flux-vacua (the problem of
generalizing mirror symmetry in the presence of fluxes was
addressed in \cite{mirror0,kstt,mirror} while for a general
analysis of compactifications on manifolds with torsion see
\cite{strominger,mirror0,berlin}).

The main purpose of the present paper is to show how theories
originating from a S-S reduction from one dimension higher, for a
suitable choice of the S-S generator, can be alternatively
interpreted as the result of a toroidal dimensional reduction in
the presence of an internal torsion. In the case of maximal S-S
supergravity in four dimensions we shall also speculate, by using
group theory techniques, on the possible M--theory flux
interpretation of (part of) the remaining  parameters of the
global symmetry group.\par
 The paper is organized
as follows:
\par In Section \ref{toy} we shall anticipate our idea with a
toy model.
\par
In  Section \ref{dimred} we shall consider a $D+n$
dimensional pure gravity theory compactified on a torus $T^n$ and
dimensionally reduced to $D$ space-time dimensions. We are going to
show
 that, by switching on an appropriate constant
background torsion in the internal $n$--torus $T^n$ (a {\em torsion
flux}), we obtain the same $D$--dimensional theory as the one
originated by a S-S reduction from $D+1$ dimensions. The S-S
generators are chosen within the global symmetry algebra
$\fsl(n-1,\R)$.
\par
In Section \ref{torsioncoup} we shall extend this
analysis to the $D=4$ maximal supergravity theory obtained from
eleven dimensions by dimensional reduction. We show that the
couplings that would appear, if performing  a S-S reduction from
$D=5$ with a $\rm SL(6,\mathbb{R})$ phase, can be interpreted in
terms of a standard dimensional reduction where the internal
manifold is given a torsion.

 In Section \ref{eleven}, we will turn to analyze the microscopic interpretation of
the $N=8,\,D=4$ supergravity obtained through a S-S
reduction from $D=5$.  Since the global symmetry group in 5
dimensions has  $78$ parameters, a general S-S phase can depend on
all of them. We shall face the problem of interpreting  these parameters in
terms of eleven dimensional background fluxes and torsion. Besides
the $36$ parameters originating from the internal torsion, and
which correspond to the choice of the S-S generator within the
$\fgl(6,\mathbb{R})$ global symmetry algebra in $D=5$, we find,
for 21 additional
 parameters, a
direct and simple interpretation in terms of the eleven
dimensional 4-form field strength, corresponding to the choice of
the S-S generator in the remaining (nilpotent) part of the Borel
subalgebra of $\fe_{6(6)}$.

In Section \ref{4dim}, we introduce the \emph{embedding tensor}
approach to gauged maximal supergravities, which is based on the
description of these models in terms of the so called T--tensor,
first introduced in the literature in Ref. \cite{dwn}. In
particular we use a four-dimensional mathematical framework in
which background
 fluxes and/or torsion are described as components of the
embedding matrix defining the corresponding gauged supergravity.
This identification, together with the characterization made in
Ref. \cite{dwst} of the embedding matrix as an $\rE_{7(7)}$
covariant tensor (transforming in the ${\bf 912}$), will allow to
describe, in maximal four dimensional supergravities originating
from Type II or M--theory, internal background
 fluxes and torsion as elements of a larger $\rE_{7(7)}$ representation.
 As a consequence of this
 mathematical characterization, in gauged lower dimensional maximal supergravities originating from
flux/torsion compactification, the duality symmetry of the
ungauged field equations and Bianchi identities can be restored
if, besides the fields, the background quantities are transformed
under duality as well. A similar analysis was made in Ref.
\cite{km} in the context of Heterotic theory compactifications.
According to our assumptions, we shall show that internal flux
and/or torsion components can indeed be consistently identified
with elements of the larger $\rE_{7(7)}$ representation in which
the embedding tensor transforms. A rule for associating flux and
torsion components with $\rE_{7(7)}$ weights is given. This
analysis provides a nice mathematical laboratory where to study
the effects of dualities on flux-- vacua (including the presence
of internal torsion as a flux of the metric moduli). As a
byproduct of this, we show that the torsion flux provides
precisely the T-dual of the NS three-form flux.

We refer the reader to the Appendix for a formal definition of
torsion and most of the technical details.

\bigskip
\section{Coupling of torsion to gauge fields:
an example} \label{toy}

 The crucial point in our investigation is the coupling of  fields
 with non vanishing spin to
gravity, in the presence of a torsion background.

Let us consider, as an example, the case of a vector field $A_M$.
The principle of general covariance demands that the  field strength
is computed in terms of covariant derivatives of the torsionfull
affine connection $\tilde\nabla$ (see the Appendix for definitions
and conventions):
\begin{equation}
F_{MN} \equiv \tilde\nabla_{[M}A_{N]}=
\partial_{[M}A_{N]}+T^P_{\ MN}A_P.
\label{torsionfull}
\end{equation}
 When $T\neq 0$ $F$ is not
invariant under the usual gauge transformations
\begin{equation}
\delta A_M(X) = \partial_M \Lambda(X).
\end{equation}
However, it is possible \cite{hrrs} to give a generalized
definition of gauge invariance which makes  the torsion compatible
with the presence of gauge fields. The generalized gauge
transformation of the field $A_M$ has  the form
\begin{equation}
\delta A_M (X)= C_M^{\ N}(X) \partial_N \Lambda(X),
\label{generalgaugetransf}
\end{equation}
where the point-dependent matrix $C_M^{\ N}(X)$ has to be
constrained by the request of gauge invariance of the field strength
\begin{equation}
\delta F_{MN}=0.
\end{equation}
 This procedure will be reviewed in section
\ref{tors}. The general result, found in Ref. \cite{hrrs}, for a
gauge field in $d$ dimensions coupled to torsion in $d$ dimensions
is\footnote{We use throughout the paper the notation $q_{(AB)} =
\frac 12 (q_{AB} + q_{BA})$ to indicate symmetrization of the
indices,
 and correspondingly  $q_{[AB]} = \frac 12 (q_{AB} - q_{BA})$ to
 indicate antisymmetrization.}
\begin{equation}C_M^{\ N} =\delta_M^N e^\phi \,, \qquad T^P_{\ MN}=
\delta^P_{[M} \partial_{N]} \phi.
\label{stringenttorsion}\end{equation} Note that the torsion tensor
in \eq{stringenttorsion} appears to be of a restricted form. There
is only one torsion degree of freedom allowed, the scalar field
$\phi$.

As we will see, by combining this idea with the dimensional
reduction procedure, we can relax the stringent conditions
(\ref{stringenttorsion}). In fact, in the spirit of the dimensional
reduction from $D+n$ to $D$ dimensions it is enough to require gauge
invariance of the $D$-dimensional theory.

The main goal of this paper will be to show that this procedure
reproduces the S-S mechanism, thus providing a geometrical
interpretation for it. This will be done in section \ref{tors}.

\bigskip

For the time being, let us illustrate our idea with a toy model,
where the metric, together with the torsion, are non dynamical,
background fields. The torsion is assumed to satisfy
(\ref{stringenttorsion}), so it is given in terms of a scalar field
$\phi$. We consider the five dimensional space-time $\R^{1,3}\times
\mathrm{S}^1$, with coordinates $x^M$, $M=0,\dots 4$, which we split
as $X^M = (x^\mu , y)$ with  $\mu = 0,...,3$  and the fifth
dimension compactified on a circle.
 For  the field $\phi$ we take
\begin{equation}
\phi = \phi(y) = m y, \qquad m=\mathrm{constant}.
\end{equation}
$\phi$ is multivalued in the circle, but the torsion is not. With
these premises we obtain
\begin{equation}
F_{MN} = \partial_{[M} A_{N]} + \partial_{[M}\phi A_{N]} =
\partial_{[M} A_{N]} - m\,\delta^4_{[M}  A_{N]},
\end{equation}
with gauge transformations
\begin{equation}
 \delta A_M =
e^{my}\partial_M \Lambda .\label{gaugey}\end{equation}
 We would like  to
make a dimensional reduction and assume $A_M(X)=A_M(x)$. But given
the gauge transformations allowed for the field $A_M$
(\ref{gaugey}), this choice cannot be preserved by a gauge
transformation. The effect of the torsion on the circle is then to
give effectively to the gauge vector a dependence on $y$. This
dependence can be chosen of the form,
\begin{equation}
A_\mu(x,y)= e^{my} A^0_\mu(x),
\end{equation}
which is preserved by gauge transformations.

 If we want to include
all the Kaluza--Klein modes, we have that the theory compactified on
the circle contains a tower of vectors
\begin{equation}
A_\mu(x,y)= e^{my}\sum_k e^{i k y}A^k_\mu (x),\end{equation}
 and
scalars
\begin{equation}
A_4(x,y)= e^{my}\sum_k e^{i k y}A^k_4 (x).\end{equation}
 We
immediately see that the effect of the internal torsion has been to
generate an exponential factor \`a la Scherk--Schwarz. Indeed, the
field strength for the zero-mode splits as
\begin{eqnarray}
F_{\mu\nu} &=& \partial_{[\mu} A_{\nu ]}, \nonumber\\
F_{\mu 4} &=& \frac 12 \left(\partial_{\mu} A_{4} - m \,
A_\mu\right) = \frac 12 e^{my}D_\mu  A^0_4,
\end{eqnarray}
where $D_\mu$ denotes a covariant derivative with respect to the
four dimensional gauge connection $A^0_\mu$. Then, the four
dimensional vector $ A^0_\mu$ gauges the translational isometry of
the action along the direction of the the axion $A^0_4$.

\section{Dimensional reduction in the presence of a torsion
background\label{dimred}}

\subsection{Dimensional reduction with a torsionless connection}

Let us consider a gravity theory in $D+n$ dimensions. We denote by
 $X^M$, $M=0,1,\cdots , D+n-1$, the coordinates in
 the space-time manifold. We will split the indices as $X^M=( x^\mu,
 y^\um )$
where the coordinates $x^\mu$, $\mu =0,1,\cdots D-1$, parametrize
non compact directions,  while $y^\um$, $\um=1,\cdots n$,
parametrize the $n$ directions of  some compact manifold
$\mathcal{M}_n$. Here and in the following, we will take
$\mathcal{M}_n= T^n$. The generalization to more general internal
manifolds will be discussed elsewhere. We will further split the
index $\um=(D, m)$ with $m=1,\dots n-1$.

The metric in the $D+n$-dimensional manifold can be conveniently
written in terms of the unconstrained fields
\begin{equation}
G_{MN}=\begin{pmatrix} e^{\frac{2\sigma}{2-D}}g_{\mu\nu} +
G_{\um\un}B^\um_\mu B^\un_\nu & \;\; G_{\um\un} B^\un_\mu
\\
 G_{\um\un} B^\um_\nu & G_{\um\un}
\end{pmatrix}
= \mathcal{V}_M^{\ A} \eta_{AB} (\mathcal{V}^T)_{\ N}^{B}.
\end{equation}
$\eta_{AB}$ is the flat metric, defined with mostly $+$ signs.
$\mathcal{V}^{\ A}=\cV_M^{\ A}dX^M $ is an orthonormal frame
(vielbein). The flat (tangent space) indices $A=1, \cdots , D+n$
split also as
$$A=(a,\ui) = (a,\hat D , i) ,\qquad a=0,\dots D-1,\quad i=1,\dots
n-1.$$ After  using partially the $\rSO(1,D+n-1)$ gauge freedom, the
vielbein can be written as
\begin{equation} \label{vielbein}
\mathcal{V}_M{}^A=\begin{pmatrix}e^{\frac{\sigma}{2-D}}\,V_\mu{}^a&V_{D}{}^{\hat{D}}\,B_\mu^D
&V_{\underline{m}}{}^i\,B_\mu^{\underline{m}}\cr 0 &
V_{D}{}^{\hat{D}} & V_{D}{}^i\cr 0&0&V_{m}{}^i
\end{pmatrix},
\end{equation}and its inverse is
\begin{equation}
\mathcal{V}_A{}^M=\begin{pmatrix}e^{-\frac{\sigma}{2-D}}\,V_a{}^\mu
& -e^{-\frac{\sigma}{2-D}}\,V_a{}^\mu\,B_\mu^D
&-e^{-\frac{\sigma}{2-D}}\,V_a{}^\mu\,B_\mu^m\cr 0 & V_{\hat{D}}{}^D
&- V_{\hat{D}}{}^D V_D{}^i\,V_i{}^m\cr 0&0&V_{i}{}^m
\end{pmatrix}\,.
\end{equation}

The vielbein of the compact manifold $\cM_n$ and its inverse are
$\rGL(n,\R)$ matrices
\begin{equation}
V_\un^{\ \ui}\equiv
\begin{pmatrix}
V_D^{\ \hat D} & V^{\ i}_D  \cr 0 & V^{\ i}_n
\end{pmatrix} \, ; \qquad  V^{\ \um}_\ui\equiv
\begin{pmatrix}
V_{\hat D}^{\ D} & V^{\ m}_{\hat D}  \cr 0 & V^{\ n}_i
\end{pmatrix}.
\end{equation}
They satisfy
\begin{equation}
 V_\um^{\ \ui} \,\eta_{\ui\uj}V_{\
\un}^{\uj}=G_{\um\un} = e^{\frac{2}{n}\sigma}
g_{\um\un}\,,\qquad\hbox{with}\quad \det\left(g_{\um\un}\right)=1.
\end{equation}
  The $\rSO(1,1)$ field $e^\sigma$ corresponds to the volume of
the internal manifold, and it has been treated separately such as to
have a canonical Einstein term in the Lagrangian in $D$ dimensions
\footnote{Indeed, with this definition one has $\sqrt{\det(G_{D+n})}
= e^{\left(1+ \frac{D}{2-D}\right)\sigma}\sqrt{\det(g_{D})} $ and
$$\mathcal{R}_{D} = R^a_{\ b \mu\nu}  \mathcal{V}_a^{\
\mu}\mathcal{V}_\rho^{\ b} G^{\rho \nu} + \cdots  \sim
e^\frac{-2\sigma}{2-D}\qquad \hbox{so that}\qquad
\sqrt{\det(G_{D+n})}\mathcal{R}_{D+n} =
\sqrt{\det(g_{D})}\mathcal{R}_{D}.$$}.

We make the Kaluza--Klein ansatz, and we truncate the Kaluza--Klein
spectrum to the 0-modes, so that
\begin{equation}\mathcal{V}_M^{\ A} = \mathcal{V}_M^{\ A}
(x).\label{ansatzvielbein}\end{equation}

 After substituting (\ref{ansatzvielbein}), the  components of
the torsionless spin connection (\ref{defconnection})
 $\omega^{AB}$ decompose as
(we use flat indices)
\begin{eqnarray}
\omega_{ab,c}&=&e^{-\frac{\sigma}{2-D}}\,\left(\frac{2}{2-D}\,\partial_{[a}
\sigma\,\eta_{b]c}+\,\bar{\omega}_{ab,c}\right)\,,\nonumber\\
\omega_{ab,\underline{i}}&=&e^{-\frac{2\sigma}{2-D}}\,\partial^{\phantom{\underline{P}}}_{[a}B^{\underline{m}}_{b]}
\,V_{\underline{m}\underline{i}}\,,\nonumber\\
\omega_{a\underline{i},b}&=&\omega_{ab,\underline{i}}\,,\nonumber\\
\omega_{a\ui,\uj}&=&
e^{-\frac{\sigma}{2-D}}\,\left(P_{\ui\uj,a}+\frac{1}{n}\eta_{\ui\uj}\,\partial_a\sigma\right)
=\omega_{a\uj,\ui}
\,,\nonumber\\
\omega_{\hat{D}i,a}&=& - \omega_{a\hat{D},i}
\,,\nonumber\\
\omega_{\hat{D}i,\uj}&=&0\,,\nonumber\\
\omega_{ij,a}&=&-e^{-\frac{\sigma}{2-D}}\,V_{[i}{}^m\,\partial_a\,V_{m|j]}\,
\equiv \,- e^{-\frac{\sigma}{2-D}} Q_{ij,a}\,,\nonumber\\
\omega_{ij,\uk}&=&0 \label{omega}
\end{eqnarray}
where we have used the definitions \begin{equation}
\partial_a \equiv V_a^\mu
\partial_\mu,\qquad e^a\equiv V^a_\mu dx^\mu ,
\end{equation}
for dual basis in the tangent and cotangent spaces, and
\begin{equation}\left(V^{-1} \, dV \right)_{\ui\uj}= \left(V^{-1} \, dV \right)_{\ui\uj,a}e^a \equiv   Q_{\ui\uj}
+P_{\ui\uj} + \frac{1}{n}\eta_{\ui\uj}d\sigma ,
\end{equation}
with
\begin{eqnarray}
&&P_{\ui\uj}=\left(V^{-1} \, dV \right)_{(\ui\uj)} -\frac
1n\eta_{\ui\uj}\rtr(V^{-1} \, dV )\,\quad \hbox{so}\quad
P_{\ui\uj}\eta^{\ui\uj}= 0,\nonumber\\
&&Q_{\ui\uj}=\left(V^{-1} \, dV \right)_{[\ui\uj]}.
\end{eqnarray}
Finally, $\bar{\omega}_{ab,c} =\bar{\omega}_{ab,\mu}V_c^{\ \mu}  $ is the torsionless spin connection of
the $D$-dimensional space-time.

\subsection{The role of torsion} \label{tors}

We want to include now the effect of a torsion background.

Let  $\Omega^A_{\ B} = \omega^A_{\ B} + \deom^A_{\ B}$ be an
antisymmetric spin-connection, with $\omega^A_{\ B}$ its torsionless
part. The torsion tensor is (see the Appendix)
\begin{equation}
T^A = d\mathcal{V}^A + \Omega^A_{\ B} \wedge
\mathcal{V}^B=\deom^A_{\ B}\wedge \mathcal{V}^B.
\end{equation}
We make the following ansatz for the dependence of the torsion on
space-time:
\begin{equation}
T^P_{\ MN}= T^A_{\ MN} \mathcal{V}_A^{\ P} = T^P_{\ MN}(X^D),
\end{equation}
 that is, we ask that the torsion tensor
depends only on the coordinate $X^D = y^1$.

As we discussed in section \ref{toy}, in the presence of torsion the
field strengths of fields with non-zero  spin  get modified as in
\eq{torsionfull} (due to the principle of general covariance)
because of the antisymmetric part in the affine connection.

Let us consider, for instance, the case of a gauge vector $A$.
 When $T\neq 0$, its field-strength $F$ is not gauge invariant.
 However, with the generalized prescription for gauge
 transformations \eq{generalgaugetransf}, that we introduced
 in section \ref{toy}, it is
possible to achieve the gauge invariance of the field strength
\begin{equation}
\delta F_{MN}=0,
\end{equation}
 thus making the torsion compatible with the presence of gauge
fields \cite{hrrs}.

On the other hand, since we would like to make the Kaluza-Klein
ansatz for all the spectrum, for our purpose it is sufficient to
assume, for the gauge parameter
\begin{equation}
\Lambda = \Lambda (x),\label{gaugeparam}
\end{equation}
 while allowing for $C_M^{\ N}$, which is associated to the presence
 of the torsion, a dependence on the
coordinate $X^D=y^1$
\begin{equation} C_M^{\ N}= C_M^{\
N}(X^D).
\end{equation}

As proven in \cite{hrrs}, the request of gauge invariance (in the
generalized sense) of the field-strength is achieved for the
generalized gauge transformation (\ref{generalgaugetransf}), under
the following condition
\begin{equation} \delta F_{MN}=C_{[N}^{\ P}
\partial_{M]}\partial_P \Lambda + \left(\partial_{[M} C_{N]}^{\ P} +
T^R_{\ MN}C_R^{\ P}\right)\partial_P \Lambda=0.\end{equation}
 With
the given ansatz \eq{gaugeparam}, this condition corresponds to the
equations
\begin{eqnarray}
&&C_{N}^{\ P} \delta^{(\mu}_{M}\delta^{\nu)}_P-C_{M}^{\ P}
\delta^{(\mu}_{N}\delta^{\nu)}_P =0 , \label{d2lambda}\\
&& \left(\partial_{[M} C_{N]}^{\ P} + T^R_{\ MN}C_R^{\
P}\right)\delta^\mu_P=0 . \label{d1lambda}
\end{eqnarray}
They may be solved for tensors $C_M^{\ N}$ and $T^P_{\ MN}$ with the
only non-zero entries
\begin{eqnarray}
C_M^{\ N}&:& \left(C_\mu^{\ \nu},C_\mu^{\ \un},C_\um^{\ \un}\right),\\
T^P_{\ MN} &:&\left(T^\mu_{\ \nu D}, T^\up_{\ MN}\right),
\end{eqnarray}
with the constraints:
\begin{equation}
C_\mu^{\ \nu}=\delta_\mu^\nu e^\phi \,; \qquad T^\mu_{\ \nu
D}=\delta^\mu_\nu\frac 12 \partial_D\phi .
\end{equation}
We note that the other non-zero entries are not restricted.

In particular,  we may consider a torsion tensor whose only non-zero
components are
\begin{equation}
T^\mu{}_{\nu D}= \frac 12 \delta^\mu_\nu\,\partial_D\phi\, ; \qquad
\,T^m{}_{n D}=-\frac 12\partial_D\Phi^m{}_n,
\end{equation}
where $\phi= \phi(X^D)$, $\Phi^m_{\ n} = \Phi^m_{\ n}(X^D)$. The
tensor $\Phi$ is taken to be a matrix of $\fsl(n-1,\R)$.

The first contribution to the torsion, $T^\mu{}_{\nu D}$, is in fact
a warping factor, which can be interpreted as an extra,
$y^1$-dependent contribution to the dilaton $\sigma$. Indeed, it
gives
\begin{equation}
d\mathcal{V}^a + \omega^a_{\ b} \wedge \mathcal{V}^b -\frac 12 d\phi
\wedge \mathcal{V}^a =0,\end{equation} that is
\begin{equation}
d\left(e^{\frac 12 \phi}\mathcal{V}^a \right) + \omega^a_{\ b}
\wedge \left(e^{\frac 12 \phi}\mathcal{V}^b
\right)=0.\end{equation} We discard such contribution (setting
$\phi =0$) because one can see that it induces non abelian
couplings for the vectors $B^m$ which are not compatible with the
definition of covariant derivatives for the scalars. It would then
introduce ghosts in the theory.

 The ansatz that we make for the non zero components of the
 torsion is:
\begin{equation}
T^m{}_{n D}=-\frac 12\partial_D\Phi^m{}_n .
\label{ansatz}\end{equation}
 We are going to show that the given
ansatz for the torsion, with  $\Phi^m_{\ n}\in \fsl(n-1)$, precisely
reproduces the Scherk--Schwarz mechanism with a phase $M\in
\fsl(n-1)\subset\fe_{n-1(n-1)}$.

Let us note that then eq. \eq{generalgaugetransf} becomes, for any
choice of tensor $C_\um^{\ \un}$,
$$\delta A_\mu = \partial_\mu \Lambda \, , \qquad \delta A_\um =0,$$
which is compatible (differently from the toy model case) with the
Kaluza--Klein ansatz $A_M= A_M (x)$.

With our ansatz, any vector field strength $F_{MN} =
\tilde\nabla_{[M}A_{N]}$ of the $D+n$-dimensional theory
decomposes into the $D$-dimensional fields:
\begin{eqnarray}
F_{\mu\nu}&=& \partial_{[\mu}A_{\nu]}, \nonumber\\
F_{\mu \un}&=&\frac 12 \partial_{\mu}A_{\un}, \nonumber\\
F_{D n}&=& \frac 12 \partial_D \Phi_{\ n}^m A_{m}.
\end{eqnarray}
Let us observe that, in order for these relations to be compatible
with the Kaluza--Klein ansatz, we have to further restrict the
torsion to be a constant:
\begin{equation}
 \partial_D \Phi_{\ n}^m =M_{\ n}^m=\mathrm{constant}\, ;\qquad
 (M_{\ m}^m=0)\,.
  \end{equation} Then, in the model that we
are considering the torsion $T^P_{\ MN}$ is a constant tensor.

\vskip 5mm In the rest of this section we will study the effects of
the torsion \eq{ansatz} on the $D$-dimensional fields coming from
the metric in $(D+n)$ dimensions.

 In the presence of torsion, the Riemann tensor has extra
contributions
\begin{equation}
\tilde R^A_{\ B} = d\Omega^A_{\ B} + \Omega^A_{\ C}\wedge
\Omega^C_{\ B} =  R^A_{\ B} + (d_{\nabla} \deom)^A_{\ B} +
\deom^A_{\ C} \wedge \deom^C_{\ B}, \label{riemann}
\end{equation}
where the symbol $d_{\nabla}$ means covariant differentiation with
respect to the torsionless part $\omega$ of the spin connection,
given in \eq{omega}.

As it is shown in the Appendix, one finds
\begin{equation}
\deom^A_{\ B |N} =  K^A_{\ B |N}, \end{equation}
 where
\begin{eqnarray}
 K^A_{\ BN}&=& \mathcal{V}_P^{\ A}
\mathcal{V}_M^{\ B} K^P_{\ MN} =\mathcal{V}_P^{\ A}
\mathcal{V}_B^{\ M}\left(
 T^P_{\ MN} -
 T_{M\phantom{P}N}^{\phantom{M}P}-T_{N\phantom{P}M}^{\phantom{N}P}
 \right),\label{kt}
 \end{eqnarray}
that is
\begin{eqnarray}
K_{AB,C}&=&(\mathcal{V}_{PA}\,\mathcal{V}_B{}^M\,\mathcal{V}_C{}^R
-\mathcal{V}_A{}^M\,\mathcal{V}_{PB}\,\mathcal{V}_C{}^R+
\mathcal{V}_A{}^R\,\mathcal{V}_B{}^M\,\mathcal{V}_{PC})\,T^P{}_{MR}\,\nonumber\\
&=&M^{\ m}_n\,(\mathcal{V}_{m [
A}\,\mathcal{V}_{B]}{}^D\,\mathcal{V}_C{}^n
-\mathcal{V}_{[A}{}^D\,\mathcal{V}_{B]}{}^m\,\mathcal{V}_{n
C}-\mathcal{V}_{m [
A}\,\mathcal{V}_{B]}{}^n\,\mathcal{V}_C{}^D)\,,\label{kt2}
\end{eqnarray}
where in the last expression we have used our ansatz \eq{ansatz} for
the torsion.
 Let us use the following short-hand notation:
\begin{eqnarray}
B_a^{\underline{m}}&=&V_a{}^\mu\, B_\mu^{\underline{m}}.
\end{eqnarray}
We get:
\begin{eqnarray}
K_{ab,c}&=&0\,,\nonumber\\
K_{ab,\hat D}&=& 0\,,\nonumber\\
K_{ab,i}&=&-e^{-\frac{2\,\sigma}{2-D}}\,M^m{}_n\,B_{[a}^D\,B^n_{b]}\,V_{mi}\,
,\nonumber\\
K_{a\hat{D},b}&=&0\,,\nonumber\\
K_{a\hat{D},\hat D}&=&0\,,\nonumber\\
K_{a\hat{D},i}&=&-\frac{1}{2}\,e^{-\frac{\sigma}{2-D}}\,
M^m{}_n\,\left(V_{\hat{D}}{}^D\,B_a^n -
B_a^D\,V_{\hat{D}}{}^n\right)\,V_{mi}\,,\nonumber\\
K_{ai,b}&=&\,K_{ab,i}\,,\nonumber\\
K_{ai,\hat D}&=&K_{a\hat{D},i}\,,\nonumber\\
K_{ai,j}&=&e^{-\frac{\sigma}{2-D}}\,M^m{}_n\,B_a^DV_{m(i}\,V_{j)}{}^n\,,\nonumber\\
K_{\hat{D}i,a}&=&-\,K_{a\hat{D},i}\,,\nonumber\\
K_{\hat{D}i,\hat D}&=& 0\,,\nonumber\\
K_{\hat{D}i,j}&=&
-\,M^m{}_n\,V_{\hat{D}}^DV_{m(i}\,V_{j)}{}^n\,,\nonumber\\
K_{ij,a}&=&e^{-\frac{\sigma}{2-D}}\,M^m{}_n\,V_{m[i}\,V_{j]}{}^n\,B_a^D\,,\nonumber\\
K_{ij,\hat
D}&=&-\,M^m{}_n\,V_{m[i}\,V_{j]}{}^n\,V_{\hat{D}}^D\,,\nonumber\\
K_{ij,k}&=& 0\,. \label{k}
\end{eqnarray}
We introduce now the structure constants $f^\um_{\ \; \un\up}$
defined as
\begin{equation} \quad f^m_{\ \;D n}= - f^m_{\
\;n D }=-M^m_{\ \, n}\,,\quad\hbox{and the rest zero},
\end{equation}
satisfying the Jacobi identities.
We can then define the non abelian field-strengths
\begin{equation}
F^\um\equiv d B^\um +\frac 12 f^\um_{\ \un
 \up} B^\un \wedge B^\up\, ,
 \label{nonabel}
 \end{equation}
or more explicitly
 \begin{eqnarray}
F^D_{ab}&\equiv& \partial_{[a}B^D_{b]},\nonumber\\
 F^m_{ab}&\equiv&\partial_{[a}B^m_{b]}+f^m_{\ D n}
 B_{[a}^D\,B^n_{b]}.
 \end{eqnarray}
Furthermore, let us define the gauge covariant derivative of a generic scalar $s_\um$ with a
covariant internal index $\um$  as
\begin{equation}
D_a \, s_\um \equiv \partial_a s_\um + f^n_{\ \um \up}B_a^\up \,
s_n\, ,
 \end{equation}
which then allows the definitions
\begin{eqnarray}
\hat P_{\ui\uj,a}&=& V_{(\ui}{}^\um \, D_a V_{\um|\uj)} -\frac
1n\eta_{\ui\uj}\,
 \rtr\left(V_{(\ui}{}^\um \, D_a V_{\um|\uj)}\right),\nonumber\\
\hat Q_{\ui\uj,a}&=& V_{[\ui}{}^\um \, D_a V_{\um|\uj]}.
\end{eqnarray}
Let us finally introduce the definitions
\begin{eqnarray}
 \,P_{ij,\hat D}&\equiv &
 M^m{}_n\,V_{\hat{D}}^DV_{m(i}\,V_{j)}{}^n\,,\qquad (P_{ij,\hat
 D}\eta^{ij}=0),\nonumber\\
Q_{ij,\hat D}&\equiv & M^m{}_n\,V_{\hat{D}}^D\,V_{m[i}\,V_{j]}{}^n
\,.
\end{eqnarray}
 From \eq{k} and
\eq{omega} it is now immediate to write down the torsionfull spin
connection, whose non-zero components  read
\begin{eqnarray}
\Omega_{ab,c}&=&\omega_{ab,c}=e^{-\frac{\sigma}{2-D}}\,\left(\bar{\omega}_{ab,c}\,+\,\frac{2}{2-D}\,\partial_{[a}
\sigma\,\eta_{b]c}\right)\,,\nonumber\\
\Omega_{ab,\ui}&=&e^{-\frac{2\,\sigma}{2-D}}\,F^\um_{ab}\,V_{\um
\ui}\,,\nonumber\\
\Omega_{a\ui,b}&=&\Omega_{ab,\ui}\,,\nonumber\\
\Omega_{a\ui,\uj}&=&e^{-\frac{\sigma}{2-D}}\,\left(\hat
P_{\ui\uj,a}+\frac{1}{n}\eta_{\ui\uj}\,\partial_a\sigma\right)
\,,\nonumber\\
\Omega_{\ui\uj,a}&=&- e^{-\frac{\sigma}{2-D}}\,\hat
Q_{ij,a}\,,\nonumber\\
\Omega_{\hat{D}i,j}&=&- \,P_{ij,\hat D}\, ,\nonumber\\
\Omega_{ij,\hat D}&=& - \,Q_{ij,\hat D}\,. \label{Omega}
\end{eqnarray}

Eq. \eq{Omega}, for $\Phi^m{}_n \in \fsl(n-1)$, may be compared with
eq. (35) in \cite{ss} with a perfect agreement.  We have then shown
that the dimensional reduction of gravity in the presence of a
torsion background of the form \eq{ansatz} is completely equivalent
to a Scherk--Schwarz model with phase $M\in \fsl(n-1)\subset
\fe_{6(6)}$.

\bigskip

Equation \eq{Omega} contains all the ingredients to write down the
$D$-dimensional Lagrangian in the presence of the torsion
background, as we are going to  see in section \ref{lagr}. The
component $\Omega_{ij}$ does not appear in the $D$-dimensional
gravity Lagrangian. However, if we consider this model as part of
the bosonic sector of a supergravity theory in $D+n$ dimensions,
it plays a role in the supersymmetrization of the model, since it
contributes to the covariant derivative of fermion fields (it is
in fact the gauged R-symmetry connection of the
($D+1$)-dimensional theory). Its components $\Omega_{ij,\hat D}$
give a mass to the gravitino and then it is responsible for the
supersymmetry breaking.

\subsection{The $D$ dimensional Lagrangian in the presence of torsion.}
\label{lagr} The $D$-dimensional gravity Lagrangian is given in
terms of the Einstein term modified by the presence of torsion. With
our conventions we have
$$\mathcal{L} =-2\,\tilde R\, \sqrt{g_{D}}\, d^D x \equiv -2
\,\tilde R^{AB}_{\phantom{AB}MN} V_A^{\ [M} V_B^{\ N]}\,
\sqrt{g_{D}}\, d^D x ,
$$
which is expressed in terms of
$$\tilde R= \partial_A \Omega^{AB}_{\phantom{AB}|B}  + \frac
12\Omega^{AC}_{\phantom{AC}|A} \Omega_{CB}^{\phantom{CB}|B} - \frac
12 \Omega^{AB|C}\Omega_{BC|A} .$$
 Let us introduce
\begin{equation}
Im \,\mathcal{N}_{\um\un} =-e^{-\frac{2\sigma}{2-D}}\, G_{\um\un},
\end{equation}
which defines  the kinetic coupling of the gauge field-strengths and
\begin{equation}
\mathfrak{V} = e^{\frac{2\sigma}{2-D}}P_{ij,\hat D} P^{ij,\hat
D}\geq 0,
\end{equation}
for the scalar potential.
It can be verified, from inspection of the above formula, that the
potential can have an absolute minimum only if $M^m{}_n \in
\mathfrak{so}(n-1)$.

\par From  \eq{Omega} we find
the $D$-dimensional lagrangian
\begin{eqnarray}
\mathcal{L}=\sqrt{g_{D}}\, d^D x\Bigl[&&-2\,\bar R
-\left(\frac{D}{(2-D)^2}+\frac 1{n}\right)\partial_a \sigma
\partial^a \sigma -
\left(\hat P_{\ui\uj|a}-\frac 1n\eta_{\ui\uj}\hat  P^\uk_{\
\uk|a}\right)^2+\nonumber\\
&&+  Im \,\mathcal{N}_{\um\un}F^\um_{ab} F^{\un |ab}-
\mathfrak{V}\Bigr],
\end{eqnarray}
where $\bar R$ is the $D$-dimensional curvature scalar.

As we are going to see in the next section, in order to obtain the
complete Lagrangian of S-S we have to fix $D=4$, $n=7$, and to
supplement the torsion flux with  non trivial  fluxes for the
four-form field-strength \footnote{Evidence for completion of the
S-S phase with the 4-form flux may also be found in Ref.
\cite{daf}}.


\section{Other couplings in maximal S-S supergravity from internal
torsion.}   \label{torsioncoup}

So far, we have been considering the effect of an internal torsion
background on the dimensional reduction of the $(D+n)$-dimensional
metric.

Let us now look at this from a slightly different point of view, in
the spirit of the toy model of Section \ref{toy}. In particular, let
us study the coupling of a gauge field $A_M$ to an internal torsion
of the form \eq{ansatz}. As discussed in section \ref{tors}, the
presence of torsion will contribute with a term of the form
\begin{equation}
F_{D m} =\frac 12 M^{n}_{\ m}A_n .
\end{equation}
The same contribution to the field strength could be obtained in the
absence of torsion by assuming an effective dependence of the vector
$A_M$ on the coordinate $X^D=y^1$, of the form (in matrix notation)
\begin{equation}
\tilde A (x,y)= U(y^1)\cdot A(x),\label{reansatz}
\end{equation}
with $U(y^1)$ given by
\begin{equation}
U(y^1)\,:\quad \begin{pmatrix} \delta^{\mu}_{\ \nu}& 0&0 \cr 0& 1
&0\cr
 0&0 & (e^{\Phi})^m_{\ n}
\end{pmatrix},\qquad \Phi^m_{\ n}=y^1M^m_{\ n}\,,\label{phase}
\end{equation}
since it would give a contribution
$$F_{Dn}=\frac 12\partial_D A (x,y)_n = \frac 12 M_{\ n}^{m}A_m.$$
 The gauge transformations that preserve the form (\ref{reansatz})
of the gauge field are of the generalized class
(\ref{generalgaugetransf}), with the tensor $C^M_{\ N}=U(y^1)^M_{\
N}$, even if,
from the $D$ dimensional point of view, the gauge transformation of
the vector $\tilde A_\mu=A_\mu$ is the usual one,
$$A_\mu\rightarrow A_\mu+\partial_\mu \Lambda(x).$$ This is in fact
the case for  any tensor $C^M_{\ N}$ which has non-trivial entries
in the components with upper index $M \neq \mu$. The matrix $U(y^1)$
is precisely of this form.

The same point of view may be applied to another field with spin,
the vielbein. We have found in section \ref{tors} that the effect of
an internal torsion background \eq{ansatz} is to induce non abelian
couplings for the four dimensional gauge vectors and scalars coming
from the metric, as in \eq{Omega}. Just as above, we observe that
the same result \eq{Omega} might have been obtained in the presence
 of a torsionless spin connection, if we had assumed
 for the
$(D+n)$-dimensional  vielbein an effective dependence on the
internal coordinate $X^D=y^1$, of the form
\begin{equation}
\tilde \cv_M^{\ A} (x,y) =U(y^1)_M^{ N}  \cv_N^{\ A} (x), \label{sstor}
\end{equation}
with  $U(y^1)$ given by \eq{phase},
which is precisely the spirit of
Scherk--Schwarz dimensional reduction (for an  $X^D$-dependent $\rSL(6)$ phase).

We observe that we can now then reinterpret  the S-S phase  as the back-reaction on the
$(D+n)$-dimensional space-time geometry of the presence of the torsion flux. \footnote{
We thank J.F. Morales for an enlightening discussion on this point.}

 \bigskip

The interplay between generalized gauge invariance in the
presence of internal torsion and S-S phases may be further studied by
considering generalized gauge invariance, in the presence of torsion,
for p-forms. Let us take, for the remaining part of this section,
$D=4$, $n=7$, and consider
 the case of the bosonic sector of eleven dimensional   supergravity.
Besides the metric $g_{MN}$,  this theory contains a three-form
$A_{MNP}$. In the presence of a torsion of type \eq{ansatz}, its
electric field strength $F_{MNPQ}$
 gets a  non
vanishing contribution due to the torsion, namely
\begin{eqnarray}
F_{Dmnp}&=& \frac 14 M^{q}_{\ [m}A_{np]q},
\label{torfluxes}
\end{eqnarray}
with $M=\partial_D\Phi \in \rSL(6,\R)$. It would then appear, in
the four dimensional theory, as an effective flux, induced by the
presence of torsion.

\bigskip

We have seen that the effect of an internal torsion on spinfull
fields may be equivalently taken into account by multiplying the
corresponding gauge field by an appropriate matrix $U^m_{\ n}(y)$
such that $T^m_{\ 4n} =(U^{-1} )^m_{\ p}\partial_y U^p_{\ n}$.
This can be understood as a consequence of the presence of
torsion, which forces the generalized rule \eq{generalgaugetransf}
for gauge invariance. In a general gauge the ($D+1$) dimensional
gauge potential, corresponding to the zero-mode in the K-K
expansion, has effectively a dependence on the coordinate $y$.
This leads to the identification of the tensor $C$, needed to
re-establish the gauge invariance, with the S-S phase $U$ as in
\eq{sstor}.

However, when considering gauge potentials corresponding to p-forms,
the generalized gauge invariance of the four dimensional
reduced theory (further restricted by the request of having also a
five dimensional interpretation) still leaves room for introducing a
more general tensor $C$ like
\begin{equation}
\delta A_{M_1\dots M_p}=  C^{N_1\dots N_q}_{M_1\dots
M_p}(y)\partial_{[N_1} \Lambda_{N_2\dots N_q]}(x),
\end{equation}
having non trivial entries only in the internal directions.

This corresponds to a freedom left to choose the fluxes in a more
general way than \eq{torfluxes}. As we are going to see in the
next section, they will complete the phase \eq{phase} to generate
a phase in the adjoint representation {\bf 78} of $\fe_{6(6)}$.
Here, appropriate fluxes, which include a generalization of
\eq{torfluxes}, provide the components needed to complete an
$\fe_{6(6)}$ phase from  the one in $\fsl(6,\R)$. The analysis
will be performed with a solvable algebra approach, in order to
make a precise identification of the four-dimensional fields in
terms of eleven-dimensional degrees of freedom.

This will complete the proof of the equivalence of the S-S model
with a flux plus torsion compactification.


\section{Completion of the S-S phase} \label{eleven}
Consider the four dimensional maximal supergravity obtained through
a S-S reduction from $D=5$. As anticipated in the introduction, the
five dimensional Lagrangian has an $\rE_{6(6)}$ global symmetry. If
this theory is thought of as originating from a toroidal
compactification of eleven dimensional supergravity, then all its
fields transform manifestly with respect to the subgroup $\rm
SL(6,\mathbb{R})\times \rm SL(2,\mathbb{R})\subset \rE_{6(6)}$,
where $\rm SL(6,\mathbb{R})$ is the group acting on the metric
moduli (except the volume) of the internal torus $T^6$, while the
dilation modulus of the $T^6$ volume, $\hat \sigma$, is contained
inside $\rSL(2,\mathbb{R})$. The five dimensional scalars
originating from the eleven dimensional fields $G_{MN}$ and
$A_{MNP}$ are:
\begin{eqnarray}
  G_{mn},\quad
A_{mnp},\quad \tilde{A}\,,
\end{eqnarray}
where $m,n,p=5,\dots, 10$;\footnote{For later convenience we adopt
here a slight change of notation, and the indices in the internal
dimensions will now run from 5 to 10.} $M,N,P=0,\dots, 10$.
$\tilde{A}$ denotes the scalar dual to
$A_{\hat{\mu}\hat{\nu}\hat{\rho}}$
($\hat{\mu},\hat{\nu},\hat{\rho}=0,\dots, 4$). All together, they
span the coset manifold ${\mathcal M}_{5}=\rE_{6(6)}/{\rm USp}(8)$,
of dimension $42=21+20+1$. $\cM_5$ can be thought of as parametrized by
the scalar fields listed above, and it  has the structure of a
solvable Lie group, inherited from the Iwasawa decomposition of the
non compact group $\rE_{6(6)}$.  We will denote by $\fsolv_5$ the
solvable Lie algebra associated to $\cM_5$ \cite{solv}. It is a
Borel subalgebra consisting of the non compact Cartan elements (in
this case all the Cartan elements can be chosen non compact)
semidirect product with a nilpotent algebra formed by the shift
generators associated to  the positive (restricted) roots. With
respect to the subgroup $\rSL(6,\R)\times \rSL(2,\R)$, the adjoint
representation ({\bf 78}) of $\rE_{6(6)}$  branches as follows:
\begin{eqnarray}
\begin{CD}{\bf 78} @>>{\rSL(6,\R)\times \rSL(2,\R)}> ({\bf 35},{\bf 1}) + ({\bf 1},{\bf 3}) + ({\bf
20},{\bf 2})\end{CD}\,, \label{sl6decomp}
\end{eqnarray}
being {\bf 35} the adjoint representation of $\rSL(6,\R)$.  Let us
denote by $t^n{}_m$ the $\fsl(6,\R)$ generators in the $({\bf
35},{\bf 1})$, $t^{\alpha\, mnp}$ the nilpotent generators in the
$({\bf 20},{\bf 2})$ ($\alpha=1,2$ being the doublet index) and by
$s^\alpha{}_\beta$ the $\fsl(2,\R)$ generators in the $({\bf 1},{\bf
3})$. A generic $\fe_{6(6)}$ generator $M$ will be a linear
combination
\begin{eqnarray}
M&=&\theta^n{}_m\,t^{m}{}_n+\xi^\beta{}_\alpha\,s^\alpha{}_\beta+\theta_{\alpha\,mnp}\,t^{\alpha\,mnp}\,.\label{M}
\end{eqnarray}
The solvable algebra $\fsolv_5$ is generated by $t^n{}_m \,\,(m\ge
n)$, $t^{1\,mnp}$ and $s^\alpha{}_\beta\,\,(\beta\ge \alpha)$. We
can choose a  coset representative $\mathbb{L}\in \rE_{6(6)}$ of
${\mathcal M}_5$ as
\begin{eqnarray}
{\mathbb L}&=&\exp{(A_{mnp}\, t^{1\,mnp})}\,\exp{(\tilde{A}\,s^1{}_2
)}\,\exp{(\sum_{m\ge n}\gamma^m{}_n\, t^n{}_m)}\,\exp{(\hat\sigma
t^1{}_1)}\,, \label{coset5}
\end{eqnarray}
$\gamma^m{}_n$ being the infinitesimal moduli defining the internal
metric (with volume normalized to 1). In fact, the vielbein of the   internal 6-torus,
$V^i=V_{m{}}^{\ i}dx^m$,
 is a  coset
representative of $\rGL(6)/\rSO(6)$, and we have
\begin{equation}V=\exp(\sum_{m\ge n}\gamma^m{}_n\,
t^n{}_m))\exp{(\hat\sigma t^1{}_1)}.\end{equation} The eleven
dimensional origin of the fields appearing in the solvable
parametrization of the scalar manifold in 5 dimensions was disclosed
in Ref. \cite{solv}.

 Consider now taking $M$ in
(\ref{M}) as a S-S generator for a S-S dimensional reduction to
$D=4$. Let us denote by $x=(x^\mu)$ the four dimensional coordinates
and by $y=X^4$ the fifth direction. Let $S$ be a generic scalar field
of the set $\{\gamma_{mn},\hat \sigma, A_{mnp}, \tilde A\}$. Then we have
\begin{eqnarray}
\mathbb{L}[S(x,y)]&=&U(y)\,\mathbb{L}[S(x)]\,h[S(x),U]\,,
\end{eqnarray}
where the compensator $h$, which is an element of $\rm USp(8)$, has
to be introduced to keep the form (\ref{coset5}). Let  $M$ be a
generator of $ \fe_{6(6)}$ of the form (\ref{M}). Since it belongs to the
isometry algebra of ${\mathcal M}_5$, it has associated a Killing
vector denoted by  $k_{(M)}$.  In terms of the solvable
parametrization, it can be written as
\begin{equation}k_{(M)}=k_{(M)}^S\partial_S.\end{equation} A  generic scalar field
$S$ in the S-S ansatz is thus expressed in the following way:
\begin{eqnarray}
S(x,y)&=&S(x)+\delta
S(x,y)=S(x)+y\,k_{(M)}^S+O(y^2)\,.\label{order1}
\end{eqnarray}
The contributions to the Lagrangian will appear in the sigma model
term  for the scalars,
\begin{equation}\langle(\L^{-1}dL)_{\mathrm{nc}},(\L^{-1}dL)_{\mathrm{nc}}\rangle ,\end{equation}
(``nc" stands for the projection on the non compact part).  One
can convince himself that the only possible contributions come for
the first order term in $y$, the one proportional to the Killing
vector himself in (\ref{order1}). This is because $k_{(M)}$ is a
global symmetry of the Lagrangian.

 The key point in order to
interpret these parameters in terms of background fluxes/tor\-sion,
is to control the higher dimensional origin of each scalar field.
This is possible thanks to the solvable Lie algebra
parametrization of the scalar manifold which we have adopted.
Indeed,  the correspondence to be considered is:
\begin{eqnarray}
\hbox{background  fluxes/torsion}\quad &\equiv &\quad
\partial_y S(x,y)_{|y=0}=k_{(M)}(x)\,,\label{correspondence}
\end{eqnarray}
being $k_{(M)}$ isometries that could in principle be gauged by the
vectors in the theory.

Let us consider a S-S generator (\ref{M})  with only the parameters
$\theta^n{}_m\neq 0$, and let us consider a particular $\L$ with
$A_{mnp}=\hat A=0$. Then the internal vielbein introduced in
\eq{sstor}  reads, in this formalism, as:
\begin{eqnarray}
V_m{}^i(x,y)&=&[U(y)\,\mathbb{L}(\gamma(x),\,\hat\sigma(x))]_m{}^i\,.
\end{eqnarray}
The parameters $\theta^n{}_m$ will enter the lower dimensional
Lagrangian through the quantity:
\begin{eqnarray}
T_{Dm}^n&=&V_i{}^n(x,y)\partial_y
V_m{}^i(x,y)_{|y=0}=\theta^n{}_m\,,
\end{eqnarray}
which coincides with the internal torsion introduced in section 3.

Similarly we can switch on only the scalar fields $A_{mnp}$,
$\tilde{A}$ and take a  S-S phase with parameters
$\theta_{1\,mnp},\,\theta^n{}_m$ and $\xi^2{}_1$. We have:
\begin{eqnarray}
A_{mnp}(x,y)&=&A_{mnp}(x)+y\,\theta_{1\,mnp}+y\,\theta^r{}_{[m}\,A_{np]r}(x)+O(y^2)\,,\nonumber\\
\tilde{A}(x,y)&=&\tilde{A}(x)+y\,\xi^2{}_1\,.
\end{eqnarray}
Using eq. (\ref{correspondence}) we can write the following
correspondence between M-theory fluxes and S-S parameters, which
generalizes \eq{torfluxes}:
\begin{eqnarray}
F_{ymnp}&=&\partial_y
A_{mnp}(x,y)_{|y=0}=\theta_{1\,mnp}+\theta^r{}_{[m}\,A_{np]r}(x)\,,\label{flux1}\\
\frac{1}{6!}\,\epsilon^{m_1\dots
m_6}\,F_{y m_1\dots
m_6}&=&\partial_y\tilde{A}(x,y)_{|y=0}=\xi^2{}_1\,.\label{flux2}
\end{eqnarray}
The second term on the right hand side of eq. (\ref{flux1}) is
required by the definition of field strength in the presence of
internal torsion. So far we have been considering the microscopic
interpretation of the S-S parameters corresponding to the choice of
M either inside the full $\fsl(6,\mathbb{R})$ or in the remaining
part of the $\fe_{6(6)}$ Borel subalgebra, namely among the
nilpotent generators $t^{1\,mnp},\,s^1{}_2$. Switching on
$\theta_{2\,mnp}$ and $\xi^2{}_1$ would introduce in the Killing
vector $k_{(M)}$, and thus on the right hand side of eqs.
(\ref{flux1}), (\ref{flux2}), more involved scalar--dependent terms.
Nevertheless, also in this case, the general eq.
(\ref{correspondence}), plus the microscopic interpretation of the
scalar fields $S$, would provide the relation between fluxes and S-S
parameters. This is work in progress, which will be presented
elsewhere.


\section{A four dimensional analysis}
In this section we shall adopt a four dimensional viewpoint and
analyze the gauging of maximal supergravity originating from a S-S
reduction using a group theoretical framework which allows to
interpret the parameters in terms of ten or eleven dimensional
quantities. This mathematical method will also prove particularly
useful in the study of the action of dualities (e.g. T-duality) on
the fields or background flux/torsion. \par The field equations and
Bianchi identities of $D=4,\,N=8$ supergravity are invariant under
the global symmetry group $\rE_{7(7)}$ \cite{cj}, which is  the
isometry group of the scalar manifold $\mathcal{M}_4=\rm
E_{7(7)}/{\rm SU}(8)$ spanned by the scalars of the theory. The
electric and magnetic charges $Q_I$ ($I=1,\dots,56$) transform in
the representation ${\bf 56}$ of $\rm E_{7(7)}$. Gauging the theory
means to promote a suitable global symmetry group $\cG\subset {\rm
E}_{7(7)}$ of the Lagrangian to be a local symmetry. To this end,
besides introducing minimal couplings which involve the vector
fields of the model, the Lagrangian should be further deformed by
the addition of fermionic mass terms and a scalar potential, which
are required by consistency of the new local symmetry with $N=8$
supersymmetry.

Let $\{t_\ell\}_{\ell=1,\cdots ,133}$ be a basis of $\fe_{7(7)}$. The most
general gauging of this theory is defined by an \emph{embedding
tensor} $\Theta_I{}^\ell$ \cite{dwst}. This tensor defines the
embedding of the gauge group $\cG$ inside $\rm E_{7(7)}$,  since it expresses the generators
$\{T_I\}$ as combinations of the $\fe_{7(7)}$ generators $t_\ell$:
\begin{eqnarray}
T_I&=& \Theta_I{}^\ell\,t_\ell\,.\label{embe}
\end{eqnarray}
In this formulation the $28 $ vector fields of the theory are
labelled by a subset of values of $I$, and since any gauging can
involve only elementary vector fields and not their magnetic
duals, $\Theta$ has necessarily rank $r\le 28$, so that the $T_I$
may be interpreted as gauge generators.
The gauged field equations and Bianchi identities are not $\rm
E_{7(7)}$--invariant anymore, since the original duality symmetry
is explicitlytly broken by $\Theta$. This global symmetry can be
formally restored if $\Theta$ is thought of as an
$E_{7(7)}$--covariant tensor in the ${\bf 56}\times {\bf 133}$ and
thus made to transform under the original duality symmetry as
well. Supersymmetry further constrains $\Theta$ to transform in
the ${\bf 912}$ of $\rm E_{7(7)}$ contained in ${\bf 56}\times
{\bf 133}$ \cite{dwst}. However different choices of $\Theta$
define different theories and therefore a transformation of
$\Theta$ should not be understood as a symmetry of the theory but
rather as a duality mapping between different gauged models. Since
compactifications in the presence of internal fluxes and/or
torsion (in the limit in which these background v.e.v. are ``small
enough'') typically give rise to a gauged lower--dimensional
supergravity, this mathematical approach is particularly suitable
for the study of dualities between different flux/torsion vacua.

In all known gauged supergravities originating from flux
compactifications, the background fluxes enter the low energy gauged
supergravity through the embedding tensor of the gauge group
\cite{flux2}. Our previous analysis, which identifies components of
an internal torsion as coupling constants in the corresponding S-S
gauged supergravity, provides strong evidence in favor of the
following correspondence:
\begin{eqnarray}
\hbox{Background fluxes/torsion}\quad&\equiv &\quad
\Theta_I{}^\ell\,.
\end{eqnarray}
The above identification tells us that background fluxes and torsion
are part of an $\rm E_{7(7)}$ representation (namely the ${\bf
912}$) and thus can be associated with suitable \emph{weights}. In
what follows we shall study this correspondence for the background
quantities which are relevant to the S-S supergravity.

\emph{ The higher dimensional origin of the four dimensional scalar
fields, vectors and coupling constants can be understood by
branching the adjoint representation of $\rm E_{7(7)}$ \footnote{The
scalar fields parametrize the manifold $\cM_4$, a solvable group
with Lie algebra  $\fsolv_4\subset \fe_{7(7)}$}, the ${\bf 56}$ and
the ${\bf 912}$ respectively, with respect to some maximal subgroup
of ${\rm E_{7(7)}}$ which characterizes the dimensional reduction.}
 Indeed, if the four
dimensional theory originates from dimensional reduction of the five
dimensional maximal supergravity, then the relevant subgroup of
${\rm E}_{7(7)}$ is ${\rm E}_{6(6)}\times \rSO(1,1)$ and we have
\cite{ssgauging0,dwst}:
\begin{eqnarray}
\fsolv_4&\longrightarrow &\fsolv_5+\fso(1,1)+\overline{{\bf
27}}_{+2}\,,\nonumber\\
{\bf 56}&\longrightarrow&\overline{{\bf 27}}_{-1}+{\bf 1}_{-3}+{\bf
27}_{+1}+{\bf 1}_{+3}\,,\nonumber\\
{\bf 912}&\longrightarrow&\overline{{\bf 351}}_{-1}+{\bf
351}_{+1}+\overline{{\bf 27}}_{-1}+{\bf 27}_{+1}+{\bf 78}_{+3}+{\bf
78}_{-3}\,,
\end{eqnarray}
where in the first branching $\fsolv_5$ is parametrized by the five
dimensional scalar fields, $\fso(1,1)$ by the radius of the fifth
dimension and $\overline{{\bf 27}}_{+2}$ by the axions $A^\Lambda_4$
originating from the 27 five dimensional vector fields
$A^\Lambda_{\hat{\mu}}$ ($\Lambda=1,\dots, 27)$. In the branching of
the ${\bf 56}$, the $\overline{{\bf 27}}_{-1}$ and ${\bf 1}_{-3}$
represent the vectors $A^\Lambda_\mu$ coming from the five
dimensional vectors and the Kaluza--Klein vector $B_\mu$, while the
remaining representations correspond to their magnetic duals. Each
representation in the branching of the ${\bf 912}$ describes a
different gauging of the four dimensional theory \cite{dwst}. In
particular the ${\bf 78}_{+3}$ defines the S-S gauging. Indeed the
corresponding embedding matrix $\Theta$ can be expressed in terms of
an element $M_\Lambda{}^\Sigma$  of ${\bf 78}$ (the S-S generator)
with grading $+3$ with respect to ${\rSO}(1,1)$ and the gauge
generators are given, from \eq{embe}, by:
\begin{eqnarray}
T_0&=&M_\Lambda{}^\Sigma\,t_\Sigma{}^\Lambda\,\,\,;\,\,\,\,T_\Lambda=M_\Lambda{}^\Sigma\,t_\Sigma
\,,
\end{eqnarray}
where $t_\Sigma{}^\Lambda$ form a basis of $\fe_{6(6)}$ while
$t_\Sigma$ are the $\fe_{7(7)}$ generators in the $\overline{{\bf
27}}_{+2}$ of $\rE_{6(6)}$. They thus close the algebra
\cite{ss,ssgauging0}
\begin{eqnarray}
\left[T_0,\,T_\Lambda\right]&=&M_\Lambda{}^\Sigma\,T_\Sigma\,;
\qquad \left[T_\Lambda,\,T_\Sigma\right] =0 \,.
\end{eqnarray}
 If the four dimensional theory is interpreted, along the
lines of the analysis followed throughout our paper, as resulting
from a two-step dimensional reduction $D=11\rightarrow
D=5\rightarrow D=4$, then we should consider branchings with respect
to the following subgroups of $\rE_{7(7)}$:
\begin{eqnarray}
{\rm SL}(6,\,\mathbb{R})\times {\rm SL}(2,\,\mathbb{R})\times {\rm
O}(1,1)&\subset &{\rm E}_{6(6)}\times \rSO(1,1)\subset
\rE_{7(7)}\,.\label{filter}
\end{eqnarray}
With respect to $\rSL(6,\,\R)\times \rSL(2,\,\R)\times \rSO(1,1)$
the S-S embedding tensor in the ${\bf 78}_{+3}$ branches as in
\eq{sl6decomp}, namely
\begin{eqnarray}
{\bf 78}_{+3} &\to & ({\bf 35},{\bf 1})_{+3} + ({\bf 1},{\bf
3})_{+3} + ({\bf 20},{\bf 2})_{+3}\,. \label{sl6decomp2}
\end{eqnarray}
Note that all the representations on the right hand side have
$\rSO(1,1)$--grading $+3$ and this is a stringent condition in
identifying the corresponding fluxes/torsion.
\par In order to associate explicit $\fe_{7(7)}$ weights to the
various fields and background fluxes and torsion,
  let us now consider the four dimensional theory as deriving
from a Type II theory in $D=10$.

We shall express the $\fe_{7(7)}$ weights in terms of a certain
orthonormal basis Cartan subalgebra
$\{\epsilon_{\underline{n}}\}$\footnote{For the sake of clarity
let us summarize the conventions on internal indices which will be
followed in the present section. Unless explicitly stated, we
shall assume $M, N=0,\dots, 10$ (labelling the eleven dimensional
space--time); $\underline{n},\underline{m}=4,\dots, 10$ (the
internal indices when reducing from $D=11$ to $D=4$, used in view
of the interpretation of $D=4$ quantities in terms of M--theory);
$n,m=5,\dots, 10$ (the internal indices when reducing from $D=11$
to $D=5$, used in view of the interpretation of $D=5$ quantities
in terms of M--theory); $r,s=4,\dots, 9$ (the internal indices
when reducing from $D=10$ to $D=4$, used in view of type IIA/IIB
interpretations of $D=4$ quantities); $u,v=5,\dots, 9$ (the
internal indices when reducing from $D=10$ to $D=5$, used in view
of type IIA/IIB interpretations of $D=5$ quantities).}
\begin{equation}\epsilon_{\underline{n}}\cdot
\epsilon_{\underline{m}}=\delta_{{\underline{n}}{\underline{m}}}\,,\qquad
\underline{n}=4,\dots 10.\end{equation} This basis is chosen in
such a way that the positive simple roots $\alpha_1,\dots \alpha_7$
are expressed as
\begin{eqnarray}
\alpha_{\underline{n}-3}&=&\epsilon_{\underline{n}}-\epsilon_{\underline{n}+1}\quad
(\hbox{for }
\underline{n}=4,\dots,8);\nonumber\\\alpha_6&=&\epsilon_8+\epsilon_9\,;
\nonumber\\
\alpha_7&=&a\quad (\mbox{IIB})\,;\qquad
\alpha_7\,=\,a+\epsilon_9\quad (\mbox{IIA})\,,\label{choices}
\end{eqnarray}
with $a$ defined as
\begin{equation}
a = -{\textstyle\frac{1}{2}} \sum_{r=4}^9\,
\epsilon_r+{\textstyle\frac{1}{\sqrt{2}}}\,\epsilon_{10} \,.
\end{equation}
The two choices (IIA) and (IIB) give isomorphic algebras
$\fe_{7,7}$, and they will be justified by the interpretation of the
fields in the higher dimensional theory.

The scalar fields parametrize the solvable group $\cM_4$, whose Lie
algebra $\fsolv_4$ is generated by the  seven (non compact) Cartan
generators of $\fe_{7(7)}$ (corresponding to the moduli $\sigma_{r}$
of the six internal radii $R_r$, $r=4,\dots, 9$, and the dilaton
$\phi$), and by $63$ shift generators associated with the positive
roots and corresponding to the axionic scalar fields. In the basis
$\{\epsilon_{\underline{n}}\}$ the correspondence between axionic
scalar fields and positive roots reads \cite{dwst}:
\begin{eqnarray}
A_{r_1r_2\dots r_k} &\leftrightarrow & a +\epsilon_{r_1}+\dots
\epsilon_{r_k} \,, \nonumber
\\
\tilde{A}_{ r_1r_2\dots r_k\mu\nu} &\leftrightarrow & a
+\epsilon_{s_1}+\dots \epsilon_{s_{6-k}}\,,\qquad
(\epsilon^{r_1\dots r_k s_1\dots s_{6-k}} \neq 0) \,,
\nonumber\\
B_{rs} &\leftrightarrow & \epsilon_{r}+\epsilon_{s} \,,
\nonumber\\
\tilde{B}_{\mu\nu} &\leftrightarrow & \sqrt{2}\,\epsilon_{10} \,,
\nonumber\\
\gamma^r{}_s&\leftrightarrow &\epsilon_{s}-\epsilon_{r}\,,\qquad (r>
s) \,, \label{wscalar}
\end{eqnarray}
By $A$ we have denoted generically the RR ten dimensional forms in
either Type IIA ($k$ even) or Type IIB  ($k$ odd) theories, and the
corresponding choices (\ref{choices}) must be understood; $B$
denotes the ten dimensional Kalb--Ramond field and $\gamma$, as
usual, are the moduli of the internal metric. The symbols with a
tilde ``\,\~\,'' denote the scalars dual to the four dimensional
2--forms.

For the vectors and their corresponding duals, since they are in the
representation ${\bf 56}$, they must be in one-to-one correspondence
with the weights $W$ of the ${\bf 56}$ of $\rE_{7(7)}$:
\begin{eqnarray}
A_{r_1\dots r_k\mu } &\leftrightarrow & w +\epsilon_{r_1}+\dots
\epsilon_{r_{k}} \,, \nonumber
\\
B_{ r\nu} &\leftrightarrow&
\epsilon_{r}-{\textstyle\frac{1}{\sqrt{2}}} \epsilon_{10}\,,
\nonumber
\\
\gamma^r_{ \mu} &\leftrightarrow&
-\epsilon_{r}-{\textstyle\frac{1}{\sqrt{2}}} \epsilon_{10}\,,
\label{wvector}
\end{eqnarray}
where
\begin{equation}
w \,=\, -{\textstyle\frac{1}{2}}\sum_{r=4}^9\,\epsilon_r \,,
\end{equation}
and the opportune choice (\ref{choices}) must be understood. The
dual potentials correspond to the opposite weights $-W$.

The above field--weight correspondence can be deduced by considering
the kinetic terms in the Lagrangian of the various fields as they arise
from a reduction of the ten dimensional theory on a straight torus.
They have the form:
\begin{eqnarray}
\mbox{dilatonic scalars:} && -\partial_\mu {h}\cdot
\partial^\mu {h} \,,
\label{hdef}
\\
\mbox{axionic scalars:} && -\frac{1}{2}\,e^{-2\,\alpha\cdot
h}\,\left(\partial_\mu \varphi\cdot  \partial^\mu \varphi \right) \,,
\\
\mbox{vector fields:} &&-\frac{1}{4}\, e^{-2\, W\cdot
h}\,F_{\mu\nu}\, F^{\mu\nu} \,,
\end{eqnarray}
where
\begin{equation}
{h}\,=\,h(\sigma,\,\phi)=\sum_{r=4}^9\sigma_r\,\epsilon_r-\sqrt{2}\,\phi_4\,\epsilon_{10}
\,,\label{h}
\end{equation}
and $\alpha$ and $W$ are the positive root and the weight associated with
the generic axion $\varphi$ and  vector field $A_\mu$ respectively.

 For the internal metric (in the ten dimensional string frame)
we have chosen $G_{rs}=e^{2\,\sigma_r}\,\delta_{rs}$ and $\phi_4$ is
the four dimensional dilaton, which is related to the ten
dimensional one by
\begin{eqnarray}
\phi_4&=&\phi-\frac{1}{2}\,\sum_{r=4}^9\,\sigma_r\,.
\end{eqnarray}
To understand the field--weight correspondence, consider for
instance the kinetic term of $A_{r_1r_2\dots r_k}$; it reads:
\begin{eqnarray}
&&\sqrt{{\rm
det}G_{(10)}}\,\,\,G_{(10)}^{\mu\nu}\,G_{(10)}^{r_1r_1}\dots
G_{(10)}^{r_k r_k}\,\partial_\mu A_{r_1r_2\dots r_k}\,\partial_\nu
A_{r_1r_2\dots r_k}=\,\nonumber\\&&\left(\sqrt{{\rm
det}g_{(4)}}\,\,\,e^{\sum_{r=4}^9\,\sigma_r+4\,\phi_4}\right)\,e^{-2\,\phi_4-2\,\sum_{i=1}^k\,\sigma_{r_i}}\,\partial_\mu
A_{r_1r_2\dots r_k}\,\partial^\mu A_{r_1r_2\dots
r_k}=\nonumber\\
&&\sqrt{{\rm det}g_{(4)}}\,\,\,e^{-2\,\alpha\cdot
h}\,(\partial_\mu A_{r_1r_2\dots r_k})^2\,,\nonumber\\&&
\end{eqnarray}
where we have used
$G_{(10)\,\mu\nu}=e^{2\,\phi_4}\,g_{(4)\,\mu\nu}$. Using the above
recipe we can associate  a weight with any flux (either of RR or
NS origin) or torsion, by inspection of the corresponding
quadratic term in the four dimensional Lagrangian. We thus
have\footnote{A similar correspondence between components of ten
dimensional field strengths and weights of the lower--dimensional
duality group was used in \cite{bill} in connection to the study
of the cosmological billiard phenomenon.}
\begin{eqnarray}
F_{\mu_1\dots \mu_p\,r_1\dots r_k}&\leftrightarrow
&-\frac{1}{2}\,\sum_{r=4}^9\epsilon_r+\sum_{i=1}^k\,\epsilon_{r_i}+
\frac{2-p}{\sqrt{2}}\,\epsilon_{10}\,,\nonumber\\
H_{r s t}&\leftrightarrow
&\epsilon_r+\epsilon_s+\epsilon_t+\frac{\epsilon_{10}}{\sqrt{2}}\,,\nonumber\\
T_{rs}^t&\leftrightarrow
&\epsilon_r+\epsilon_s-\epsilon_t+\frac{\epsilon_{10}}{\sqrt{2}}\,,\label{wflux}
\end{eqnarray}
where $F=dA$, $H=dB$ are the RR and NS field-strengths and $T$ is
the internal torsion.

\bigskip

After this formal treatment, we can now make contact with the
analysis made in the previous section about S-S reduction from
five dimensions. We had interpreted the five dimensional scalar
fields from an 11 dimensional (M--theory) point of view. They correspond to
$A_{mnp},\,\tilde{A}$ and $\gamma^n{}_m$. When we perform a
reduction from 11 to 10 dimensions, and then to 4, they can be
interpreted in a Type IIA language as
$A_{uvw},\,B_{uv}=A_{uv\,10},\,\tilde{A}$,
$\gamma^u{}_v\,\,(u>v),\,A_u=\gamma^{10}{}_v$, where
$u,v,w=5,\dots, 9$ and the corresponding positive roots are
(\ref{wscalar}):
\begin{eqnarray}
A_{uvw}&\leftrightarrow & a
+\epsilon_u+\epsilon_v+\epsilon_w\,\,,\nonumber\\
\tilde{A}&\leftrightarrow&
a+\sum_{r=5}^9\,\epsilon_r\,,\nonumber\\
A_u&\leftrightarrow& a
+\epsilon_u\,,\nonumber\\
B_{uv}&\leftrightarrow &\epsilon_u+\epsilon_v\,,\nonumber
\\
\gamma^u{}_v&\leftrightarrow& - \epsilon_u+\epsilon_v\,.
\end{eqnarray}
 The $\rSO(1,1)$ parametrizing
the radius of the fifth ($X^4$) dimension is generated by the Cartan
generator $H_\lambda$ associated to the weight
$\lambda=2\,\epsilon_4+\sqrt{2}\,\epsilon_{10}$ (which is the
highest weight of the ${\bf 56}$ of $\rE_{7(7)}$).
 The M--theory fluxes which we have associated with (part of the)
 S-S parameters
are $F_{4mnp}$ and $F_{4m_1\dots m_6}$. They are interpreted from
Type IIA point of view as the fluxes
$F_{4uvw},\,H_{4uv}=F_{4uv\,10}$ and $\tilde{F}_{\mu\nu\rho\sigma}$
and are associated with the weights:
\begin{eqnarray}
F_{4uvw}&\leftrightarrow &
-\frac{1}{2}\,\sum_{r=4}^9\epsilon_r+\epsilon_4+\epsilon_u+\epsilon_v+\epsilon_w+\sqrt{2}\,\epsilon_{10}\,,\nonumber\\
H_{4uv}&\leftrightarrow
&\epsilon_4+\epsilon_u+\epsilon_v+\frac{\epsilon_{10}}{\sqrt{2}}\,,\nonumber\\
\tilde{F}_{\mu\nu\rho\sigma}&\leftrightarrow
&\frac{1}{2}\,\sum_{r=4}^9\epsilon_r+\sqrt{2}\,\epsilon_{10}\,,\label{ssf1}
\end{eqnarray}
where $\tilde{F}_{\mu\nu\rho\sigma}$ is the 7-form flux dual in
eleven dimensions to $F_{\mu\nu\rho\sigma}$ and its weight is thus
the opposite to the corresponding one which can be read off from
eqs. (\ref{wflux}). As far as the internal torsion $T_{4n}^m$ is
concerned, it corresponds, from the Type IIA point of view, to the
ten-dimensional internal torsion $T_{4u}^v$ and to a component of the RR two-form field strength,
$F_{4u}=T_{4u}^{10}$. They are associated with the weights:
\begin{eqnarray}
T_{4u}^v&\leftrightarrow &
\epsilon_4+\epsilon_u-\epsilon_v+\frac{\epsilon_{10}}{\sqrt{2}}\,,\nonumber\\
F_{4u}&\leftrightarrow &
-\frac{1}{2}\,\sum_{r=4}^9\epsilon_r+\epsilon_4+\epsilon_u+\sqrt{2}\,\epsilon_{10}\,.\label{ssf2}
\end{eqnarray}
Using the weight representation \eq{ssf1}, \eq{ssf2} of the relevant
fluxes and torsion involved in the S-S gauging, one may check that
their identification with components of the S-S embedding tensor in
the ${\bf 78}_{+3}$ is indeed consistent. Here we shall only give
some evidence based on grading arguments. The $\rSO(1,1)$ weight
associated with the above fluxes is given by the scalar product of
$\lambda$ times the related weight. As it can be easily verified
this product gives $+3$ for all the above fluxes. Indeed these
weights can all be expressed as
\begin{equation}
W_{\mbox{(S-S flux/torsion)}}=\lambda/2+\beta\,\end{equation}
where $\beta$ is a root of $\fe_{6(6)}$ such that $\lambda\cdot
\beta=0$.

 This is consistent with our interpretation of these
flux/torsion components as belonging to the ${\bf 78}_{+3}$.

 Let us
now comment on the transformation properties of the fluxes/torsion
in \eq{ssf1}, \eq{ssf2} with respect to the ${\rm SL}(2,\mathbb{R})$
group in \eq{filter}. We find that the fluxes $F_{4uvw},\,H_{4uv}$
belong to a doublet (the ${\bf(20,2)}_{+3}$ in \eq{sl6decomp2}),
$\tilde{F}_{\mu\nu\rho\sigma}$ to a triplet (the ${\bf(1,3)}_{+3}$
in \eq{sl6decomp2}) while $T_{4u}^v\,\,(v>u),\,F_{4u}$ are singlets
(in the ${\bf(35,1)}_{+3}$ in \eq{sl6decomp2}). To show this it
suffices to check the grading relative to the Cartan generator
$H_{\tilde\alpha}$ of the corresponding $\fsl(2,\mathbb{R})$ algebra. The
positive root $\tilde\alpha$ of $\fsl(2,\R)$  is
\begin{eqnarray}
 \tilde\alpha=a+\sum_{r=5}^9\epsilon_r\,,
\end{eqnarray}
and it is the one corresponding to the scalar $\tilde{A}_{4\mu\nu}$.
The $H_{\tilde\alpha}$ grading of any field or flux/torsion is given by the
scalar product of $\tilde\alpha$ by the corresponding weight. It is
normalized  $+1$ for the highest weight component of a doublet, $+2$
for the highest weight component of a triplet and $0$ for a singlet.
One can easily check that:
\begin{eqnarray}
\tilde{F}_{\mu\nu\rho\sigma}&:&\qquad \tilde\alpha\cdot
W=+2\,,\nonumber\\
F_{4uvw},\,H_{4uv}&:&\qquad \tilde\alpha\cdot W=+1\,,\nonumber\\
T_{4u}^v,\,F_{4u}&:&\qquad \tilde\alpha\cdot W=0\,.\nonumber
\end{eqnarray}
This proves the above statement.
\paragraph{Fluxes, torsion and T--duality.\label{4dim}}
An important advantage of the above mathematical setting is that we
can make a simple characterization of the action of T--duality on
either fields or fluxes/torsion \cite{bt}. Let us consider for the
sake of simplicity  the dimensional reduction of a Type II theory on
a straight torus. The effect of T--duality along the internal
direction $X^r$ is to transform the corresponding radius $R_r$ and
the dilaton $\phi$ as follows ($\alpha^\prime=1$) \cite{polch}:
\begin{eqnarray}
R_r&\rightarrow& \frac{1}{R_r}\,\,\Rightarrow
\,\,\,\sigma_r\rightarrow \sigma_r^\prime=-\sigma_r\,,\nonumber\\
\phi&\rightarrow &\phi^\prime=\phi-\sigma_r\,\qquad r=4,\dots 9.
\end{eqnarray}
As a result the $h(\sigma,\,\phi)$ vector, defined in \eq{h},
transforms into a new vector
$h^\prime=h(\sigma^\prime,\,\phi^\prime)$ and this in turn can be
absorbed in a transformation of the various $\rE_{7(7)}$ weights
which contract $h$ in the reduced Lagrangian: $W\cdot
h(\sigma^\prime,\,\phi^\prime)=W^\prime \cdot h(\sigma,\phi)$.
Therefore the effect of T--duality along a number of directions
$X^r$ is to change the sign of $\epsilon_r$ inside the expression of
$\rE_{7(7)}$ weights:
\begin{eqnarray}
\mbox{T--duality along $X^r$}&\Rightarrow & \epsilon_r\rightarrow
-\epsilon_r\,.\label{te}
\end{eqnarray}
This action is consistent with the characterization of T--duality as
an \emph{automorphism} of the group ${\rm SO}(6,6)$ which acts on
the $\gamma^r{}_s,\,B_{rs}$ moduli \cite{stelle,bt}. In particular a
T--duality transformation along an odd number of internal
directions, which maps Type IIA and Type IIB theories into each
other, corresponds to an \emph{outer} automorphism of the algebra
${\fso}(6,6)$.

 Given the description of fields, fluxes and
torsion in terms of $\rE_{7(7)}$ weights as in eqs. (\ref{wscalar}),
(\ref{wvector}), (\ref{wflux}), we can now establish (at a
linearized level) the action on them of  T--duality.
 From this
framework it follows naturally (see \eq{wflux}) that the flux $H_{rst}$ and the
torsion $T_{rs}^t$ are mapped into each other by a T--duality along
the direction $X^t$:
\begin{eqnarray}
\mbox{T-duality along $X^t$:}&&\,\,\,H_{rst}\longrightarrow
T_{rs}^t\,.
\end{eqnarray}
This motivates, at the level of maximal supergravity or of
truncations of it\footnote{For example in the toroidal orientifold
models studied in \cite{flux3,IIAflux}.}, the duality correspondence
between a vacuum with H--flux in Type II B(A) and a background with
internal torsion in Type II A(B).\par
 It should be stressed that
the above treatment does not immediately give an interpretation in
terms of flux or torsion of all the weights of ${\bf 912}$.


\section{Conclusions}
In this paper we have studied the interpretation of the coupling parameters
entering the four-dimensional $N=8$ S-S supergravity in terms
of internal background torsion and form--fluxes in a dimensional
reduction from ten or eleven dimensions. Some of these parameters
have an immediate interpretation from a higher dimensional point
of view, for some of the others this interpretation is more subtle
as they correspond to non--perturbative symmetries of the
microscopic theory.\par There are various directions for future
investigations, which include a similar analysis of theories
corresponding to compactifications on more general manifolds with
reduced holonomy, the construction of a $D=4$ gauged supergravity
deriving from a general torsion background and the effects of a
dynamical torsion on this setting. Finally it would be interesting
to use our analysis to develop a solution generating technique, in
the spirit of \cite{lm}, which uses the full $\rE_{7(7)}$ duality
group to connect different supergravity solutions.


\appendix
\section{Some useful relations and definitions about the torsion}

We consider a Riemannian manifold with metric $g_{MN}$. We want to
consider an affine connection $\Gamma$ on this manifold, and we
require the compatibility of the connection with the metric,
\begin{equation}\nabla_P\,g_{MN}=0.\label{metricompatible}\end{equation}
If we assume the torsionless condition $ \Gamma_{MN}^P=
\Gamma_{NM}^P$, this equation determines, as unique solution, the
Levi--Civita connection, given by
\begin{equation}(\nabla_MX)^P=\partial_MX^P+\Gamma_{MR}^PX^R,\qquad
\Gamma_{MN}^P=\frac
12g^{PR}(g_{RM,N}+g_{RN,M}-g_{MN,R}),\label{levicivita}\end{equation}
for the covariant derivative of a vector field $X$.

The torsionlessness condition can be relaxed so the connection
coefficients $\Gamma$ for $\nabla$ are not uniquely determined in
general. Nevertheless, if we insist on the request of a metric
connection, equation (\ref{metricompatible}) determines the
connection coefficients in terms of the torsion tensor
$$T_{\;\;MN}^P=\tilde\Gamma_{[MN]}^P=\frac 12
(\tilde\Gamma_{MN}^P-\tilde\Gamma_{NM}^P)\,,$$ as
\begin{equation}\tilde\Gamma_{MN}^P=\Gamma_{MN}^P+\frac 12(T_{\
MN}^P-T_{M\ N}^{\ P}-T_{N\ M}^{\
P}),\label{connection}\end{equation} where $\Gamma_{MN}^P$ are the
coefficients of the Levi-Civita connection. The indices are raised
and lowered with the metric. We notice that the two last terms in
(\ref{connection}) give a contribution to the symmetric part of the
connection. We will denote
$$K^P_{\phantom{P} MN}=\frac 12(T_{\phantom{P}
MN}^P-T _{M\phantom{P} N}^{\phantom{P} P}-T_{N\phantom{P}
M}^{\phantom{P} P}).$$ Note that $K^P_{\ MN}$ has not definite
symmetry in its last two indices, but it is instead antisymmetric in
its first two: $$K^P_{\phantom{P} M N} = -
K^{\phantom{M}P}_{M\phantom{P} N}.$$

\bigskip

\bigskip

We will use the formulation of orthonormal frames. The affine
connection, defined by $\tilde \Gamma^P_{MN}$, transforms from the
curved frame to the orthonormal one as
\begin{equation}\Omega_{\phantom{P} A|M}^B=
\tilde\Gamma_{\phantom{P}
MN}^R\cV^N_A\cV^B_R+\cV^B_N\partial_M\cV^N_A,\label{changebasis}\end{equation}
where  $\Omega_{\
A}^B=\Omega_{\phantom{P} A|M}^BdX^M$ is the {\it spin connection}.
If $\tilde\Gamma$ is a metric
 connection, then the spin connection satisfies
 $$\Omega_{AB}=-\Omega_{BA},$$
 where the flat indices are raised and lowered by the flat metric
 $\eta_{AB}$.
 The
 torsion two form is the covariant differential of the identity $\id
 =\partial_M\otimes
 dx^M$, so
 $$T^M=d_{\tilde \nabla}(dX^M)\equiv ddX^M+\tilde \Gamma^M_N\wedge
 dX^N=\tilde\Gamma_{\ RN}^MdX^R\wedge dX^N.$$
 In the orthonormal frame we have $\partial_M=\cV^A_Me_A$, where
 $e_A$ are the orthonormal
 vectors. In this basis
 the torsion tensor is
$$T^A=d \cV^A+ \Omega^A_{\ B}\wedge \cV^B.$$
If $T^A=0$ the spin connection  is related through
(\ref{changebasis}) to the Levi--Civita connection
(\ref{levicivita}). It can be written in terms of the vielbein as
\begin{equation}\omega^{AB}_M = \cv_C^{\ N} \cv_{D}^{\ R}
\eta^{AD}\eta^{BC} \left(q_{R|MN} - q_{M|NR} +
q_{N|RM}\right)\,,\label{omq}
\end{equation}
with
\begin{equation}
q_{R|MN} \equiv \eta_{AB} \cv_R^{\ A} \partial_{[M}\cv_{N]}^{\
B}\,,
\end{equation}
(for a derivation, see, e.g. \cite{cast}). We denote the difference
between $\Omega$ and $\omega$ as $\deom$. Since the difference
between two connections is a tensor we have
$$
\deom^A_{\ B |M}=\Omega^A_{\ B |M}-\omega^{AB}_M=\mathcal{V}_B^{\ N}
K^P_{\phantom{P}NM} \mathcal{V}_P^{\ A} = K^A_{\ B |M},
$$
and it is related to the torsion form as
$$T^A=d\cv^A + \Omega^A_{\ B }\wedge \cv^B = \deom^A_{\ B |M} \cv^{\
B}_N dX^M \wedge dX^N = K^A_{\phantom{P}[NM]}  dX^M \wedge dX^N .$$

\subsection{Decomposition of $\omega^A_{\ B}$ in terms of
$D$-dimensional fields}

Equation (\ref{omq}) can be written in the following
equivalent form:
\begin{eqnarray}
\omega_{AB,C}&=&q_{A|CB}+q_{C|AB}+q_{B|AC}\,,\nonumber\\
q_{C|AB}&=&\mathcal{V}_{[A}{}^M\,\mathcal{V}_{B]}{}^N\,\partial_{M}\mathcal{V}_{NC}\,\nonumber\\
&=&q_{P|MN} \cv_C^{\ P} \cv_{[A}^{\ M} \cv_{B]}^{\ N}.
\label{defconnection}
\end{eqnarray}

In our case, given \eq{vielbein}, with all the corresponding fields
only depending on $x^\mu$, we have
\begin{eqnarray}
q_{R|\um\un}&=& 0\,, \nonumber\\
q_{\un| \mu\um}&=& \frac 12 \eta_{\ui\uj}\cv_\un^{\ \ui}
\partial_\mu \cv_\um^{\ \uj}\,,\nonumber\\
q_{\nu| \mu\um}&=& \frac 12 \eta_{\ui\uj}\cv_\nu^{\ \ui}
\partial_\mu \cv_\um^{\ \uj}\,,\nonumber\\
q_{\um| \mu\nu}&=&  \eta_{\ui\uj}\cv_\um^{\ \ui} \partial_{[\mu}
\cv_{\nu]}^{\ \uj}\,,\nonumber\\
q_{\rho| \mu\nu}&=&  \eta_{\ui\uj}\cv_\rho^{\ \ui}\partial_{[\mu}
\cv_{\nu]}^{\ \uj} + \eta_{ab}\cv_\rho^{\ a}\partial_{[\mu}
\cv_{\nu]}^{\ b}.
\end{eqnarray}

\subsection{D-dimensional Kaluza--Klein vectors}
Consider a diffeomorphism on the $x^\um$ coordinates which is local
with respect to the D-dimensional space-time: $x^D\rightarrow
x^\um+\xi^\um (x^\mu)$. Under such transformation, with the
Kaluza--Klein assumption, we have:
\begin{eqnarray}
\delta \mathcal{V}_\mu^{\ A}&=& \partial_\mu \xi^\um
\mathcal{V}_\um^{\ A}\nonumber\\
\delta \mathcal{V}_\um^{\ A}&=&0
\end{eqnarray}
which gives
\begin{equation}
\delta B^\um_\mu= \partial_\mu\xi^\um \,,\\
\end{equation}
The diffeomorphisms in the $y^\um$ directions appear therefore as
gauge transformations for the abelian vector $B^\um_\mu$, and, in
absence of torsion, the $B^\um$ behave as good  abelian vectors in
$D$ dimensions.

\section*{Acknowledgements}

We are thankful to R. D'Auria and S. Ferrara for reading a
preliminary draft of our paper and to G. Bonelli and J. F. Morales
for enlightening discussions.

This work has been supported in part by the European Community's
Program  contract MRTN-CT-2004-005104.

The work of M. A. Ll. has been supported in part by the research
grants BFM 2002-03681 from the Ministerio de Ciencia y
Tecnolog\'{\i}a (Spain) and from EU FEDER funds, and by the research
grant GV04B-226 rom the Generalitat Valenciana.

L.A. would like to thank the Physics department of the Politecnico
di Torino, the PH-TH-division of CERN and the Departamento de
F\'{\i}sica Te\'orica, Universidad de Valencia,  for their kind
hospitality during the preparation of the manuscript.

M. A. Ll. would like to thank the PH-TH-division of CERN for its
kind hospitality during the preparation of this work.



\end{document}